\documentclass[letterpaper]{article} %
\usepackage{times}  %
\usepackage{aaai2026} 
\usepackage{helvet}  %
\usepackage{courier}  %
\usepackage[hyphens]{url}  %
\usepackage{graphicx} %
\usepackage{natbib}  %
\usepackage{caption} %
\usepackage{algorithm}
\usepackage{algorithmic}

\usepackage{newfloat}
\usepackage{listings}
\DeclareCaptionStyle{ruled}{labelfont=normalfont,labelsep=colon,strut=off} %
\floatstyle{ruled}
\newfloat{listing}{tb}{lst}{}
\floatname{listing}{Listing}
\title{Identifying Harm in Personalized, Generative AI Systems Requires User-Centered Auditing at the Interaction Level}
\author {
    Hannah Cha
}
\affiliations{
    Microsoft Research\\
    Stanford University\\
   v-hannahcha@microsoft.com
}

\begin{document}

\maketitle

\begin{abstract}
Personalized, generative AI systems increasingly adapt their behavior to individual users over time, fundamentally changing model behavior. While existing auditing approaches have been effective at surfacing harms in non-personalized contexts, they often rely on static, simulated evaluations and definitions of harm that aggregate across broad, group categories. In this position paper, we argue that such approaches can fail to capture emergent harms in personalized generative AI systems, where harms surface through interpretations of ongoing interaction and evolve with user history. We identify three presuppositions underlying many harm auditing paradigms: that harms can be (1) specified outside real-world interaction, (2) defined non-pluralistically within groups, and (3) treated as static. One might argue that personalized systems could simply learn definitions of what constitutes harm to individual users through repeated interactions. However, we argue that attempts to surface user harms through deeper personalization risk imposing asymmetric burdens of labor and privacy on marginalized users. Consequently, we propose reframing understandings of harm as adaptive, user- and community-centered processes, and outline design directions that shift auditing from retrospective evaluation toward infrastructures that support ongoing articulation of harm in interaction. Our work highlights the need for auditing and design practices that better reflect the pluralistic and evolving nature of harm understanding in personalized generative AI systems.
\end{abstract}

\section{Introduction}
Generative AI systems are being increasingly integrated into everyday applications, such as chatbots~\citep{neumann2024llm, dam2024complete} and writing assistants~\citep{mysore2024pearl, guo2024understanding}. These systems have increasingly grown to personalize, or adapt outputs to individual users~\cite{eapen2023personalization}, based on past behavior~\citep{ning2025user, xu2025personalized} and user preferences~\citep{kang2023llms, gao2024aligning}. These methods promise more relevant and engaging interactions for users~\citep{kim2025cupid, ha2024clochat}. While personalization is not new in computational systems~\citep{lu2012recommender, joachims2002optimizing, narayanan2004adaptive}, the nature of personalization in generative AI is fundamentally different. Unlike selection-based personalization, which chooses from a fixed set of outputs, generative AI reshapes the distribution of possible outputs, producing indeterminate, context-dependent, and evolving behaviors~\citep{ning2025user, kowsher2024token}. In fact, personalization can fundamentally alter the nature of model behavior itself~\citep{WANG2025101397}. As personalization continues to shape the ways that systems interact with users, it is crucial to be able to document exactly when and how harms may emerge.

Existing paradigms to audit harms in generative AI systems often occur as non-personalized, simulated evaluations~\citep{sandvig2014auditing, hofmann2024ai, salinas2024s}, where the model does not retain memory of any prior interactions and the simulated context can be abstracted away from real world use. While these approaches have systematically revealed biases in baseline models~\citep{hofmann2024ai, salinas2024s, mittermaier2023bias, lewis2025unpacking}, they may fail to capture true user harms arising as a result of personalized interaction. Harm cannot be determined solely by what the output constitutes, but from the way individual users experience and interpret these outputs~\citep{wang2025measuring, brancaleone2025within, mosley2025social}. As AI systems become personalized to individual users, merely auditing for harms in non-personalized contexts risks missing emergent harms that impact real users. Additionally, current systems that users interact with provide few mechanisms for users to articulate harm, contest system assumptions, or directly shape how personalization unfolds over time. In the absence of such mechanisms, personalized AI systems default to making inferences about users~\citep{purificato2024usermodelinguserprofiling}. Ultimately, this risks treating user preferences, such as what constitutes harm, as something that can be inferred from proxies~\citep{WANG2025101397} rather than directly articulated by users.

In this position paper, we specifically focus on \emph{experiential harms}, which consist of harms that users can self-identify and report experiencing negative affect from~\citep{wang2025measuring}. We argue that personalized generative AI systems create experiential harms during the course of \emph{interaction} that are often difficult to surface through traditional auditing mechanisms because interactions can create emergent behavior. More specifically, we argue that many, general-purpose auditing approaches presuppose that: 
\begin{enumerate}
    \item \textbf{Harms can be specified outside of real-world interaction}: Auditing often happens in a memoryless, non-personalized contexts, abstracted away from how systems may truly harm users in the real world.
    \item \textbf{Harms can be defined non-pluralistically within groups}: Auditing often defines harm at broad, demographic categories (e.g. race, gender), which can obscure the pluralistic, intersectional ways members of the same group can experience harm.
    \item \textbf{Harms are static}: Many auditing methods lack continual or evolving evaluation over time, failing to capture how personalization introduces new harmful behavior over time, or how user definitions of harm may evolve.
\end{enumerate} 
We further argue that attempting to personalize definitions of harm to individual users with current system paradigms can create asymmetries of \emph{labor} and \emph{privacy}, disproportionately burdening users in minoritized groups. To address these challenges and bridge the gap in understanding true user harms, we propose directions for auditing that acknowledge defining harm as an \emph{adaptive, user- and community-centered process}, emphasizing evolving, pluralistic understandings of harm without causing disproportionate burden on users in minority groups. Ultimately, this work motivates the need for better harm auditing mechanisms that allow all users, especially those from minoritized backgrounds, to reap the benefits of personalized, generative technologies.

\section{Related Work}
\subsection{Personalized, Generative AI Systems}
Personalization is not new in computational systems. Recommendation systems~\citep{lu2012recommender}, search engines~\citep{joachims2002optimizing}, and adaptive interfaces~\cite{narayanan2004adaptive} have long tailored outputs to users. In these traditional systems, personalization primarily affects selection: which item is recommended~\citep{ricci2010introduction}, which result is ranked higher~\citep{joachims2002optimizing}, or which predefined label is assigned~\citep{barocas2016big}. Although users may interpret these outputs differently, the system’s behavior itself often remains stable and enumerable. Thus, algorithmic decision rules can be probed to surface discrimination or harms because outputs are repeatable and enumerable across test inputs~\citep{sandvig2014auditing}. 

By contrast, generative AI systems produce outputs through stochastic generation that may be conditioned on user behavior and preferences~\citep{ning2025user, xu2025personalized}, prior conversational history~\citep{kowsher2024token}, or even inferred attributes~\citep{salinas2024s}. Thus, personalization in generative settings does not merely select among existing outputs, but reshapes the distribution over possible behaviors. The system may adopt different tones, frames, assumptions, and normative stances depending on the user it is interacting with~\citep{wang2025personalization}. In fact, personalization fundamentally changes model behavior, even modifying behavior on standard benchmarks~\citep{WANG2025101397}. Consequently, the model does not exhibit a single, stable behavior, but rather a family of behaviors that continually evolves in a personalization context that requires new ways of evaluation~\citep{hopkins2025chatbot}. 

Recent work has attempted to better understand and evaluate how personalized interaction can affect behaviors for generative AI systems, such as chatbots~\citep{geng2025accumulating, shen2024position, wang2025personalization}. We extend this need for new evaluation paradigms for personalized, generative AI systems specifically for the harms that users may experience.

\subsection{Surfacing Harms Via Algorithm Auditing}
Algorithm auditing describes the process of surfacing harms by directly analyzing the outputs of such systems~\citep{birhane2024aiauditingbrokenbus, lacmanovic2025artificial, HCI-083}. Such harms include stereotypes~\citep{ghosh2024generative}, problematic outputs~\citep{birhane2024aiauditingbrokenbus, lacmanovic2025artificial}, or surfacing demographic-based disparities in outcomes in domains such as healthcare~\citep{obermeyer2019dissecting, mittermaier2023bias, seyyed2021underdiagnosis}, employment~\citep{chen2018investigating, wilson2024gender, wen2025faire}, and housing~\citep{Asplund_Eslami_Sundaram_Sandvig_Karahalios_2020, zou2023ai, juhn2022assessing}. Many of these audit-based methods, especially in the context of generative AI systems, often consist of researchers simulating a use case of an LLM through prompts and iteratively evaluating a series of AI outputs. Potential harms from these systems are often surfaced by providing proxies for demographic groups, such as name~\citep{salinas2024s} or dialect~\citep{hofmann2024ai}, and measuring discrepancies across groups. Such simulated studies have measured these discrepancies in the natural language used to describe groups~\citep{hofmann2024ai, lim2024african, cheng2023marked}, outcomes groups achieve~\citep{salinas2024s, bouguettaya2025racial}, or the way marginalized groups are portrayed in generative imagery~\citep{wang2025large, yang2025racial}. 

However, one common critique of these audit-based methods is that these evaluations are created and run by either AI practitioners~\citep{HCI-083, sandvig2014auditing, sweeney2013discrimination} or AI researchers~\citep{buolamwini2018gender, cramer2018assessing, eslami2019user, eslami2017careful, raji2020closing}, who may be disconnected from the communities actually experiencing harm~\citep{blodgett2020language}. As a result, there has been a rise in end users and communities identifying emergent harms from AI systems~\citep{devos2022toward, shen2021everyday}. Users can discover nuanced and community-specific harms that have previously been overlooked by researchers or practitioners ~\cite{mack2024they, mim2024between, zhang2024partiality,young2019toward, qadri2025casethickevaluationscultural}. For instance, ~\citet{deng2025weauditscaffoldinguserauditors}'s WeAudit created an infrastructure allowing for users to audit amongst community members and report harms to AI practitioners in actionable ways. These user-centered auditing practices have been helpful in surfacing community-specific harms, such as stereotypical representations of non-Western cultures~\citep{ghosh2024generative}, inappropriate outputs in cultural contexts~\citep{qadri2025casethickevaluationscultural} and the lack of disability~\citep{mack2024they} and gender~\citep{ghosh2024don} representation in generative image models. Overall, these auditing methods focus on experiential harms, or harms that can be surfaced through self-reported experiences of negative affect~\citep{wang2025measuring}, from the users themselves. 

Nonetheless, existing auditing approaches can miss harms emerging from the ways that users interact with personalized AI tools. Simulated, stateless audits happen outside of true user interaction, and many community-based methods may consult communities at the design phase rather than continuing to evaluate the impact of tools after they are deployed~\citep{harrington2019deconstructing}. Furthermore, even end-user auditing often functions with the goal of providing technical recommendations and remediations for AI practitioners~\cite{deng2025weauditscaffoldinguserauditors, saleiro2019aequitasbiasfairnessaudit,raji2020closing, mokander2023auditing}, rather than preventing harm or reshaping outputs for users as they occur. This can lead to an asymmetry of burden and benefit, where those who perform the work required by the system may not be the one who receive its benefits~\citep{grudin1988cscw}. Ultimately, our work highlights this gap in auditing paradigms by outlining the nature of harm in personalized, generative AI systems that current auditing methods may fail to capture. 

\section{Why Existing Auditing Approaches Are Insufficient for Personalized Generative AI Systems}
In this section, we argue that personalized, generative AI systems create harms that current auditing approaches may be insufficient for surfacing. We outline how existing auditing approaches often presuppose harm (1) can be specified outside of real world interaction, (2) is non-pluralistic within groups, and (3) is static.

\begin{figure*}
\includegraphics[width=\linewidth]{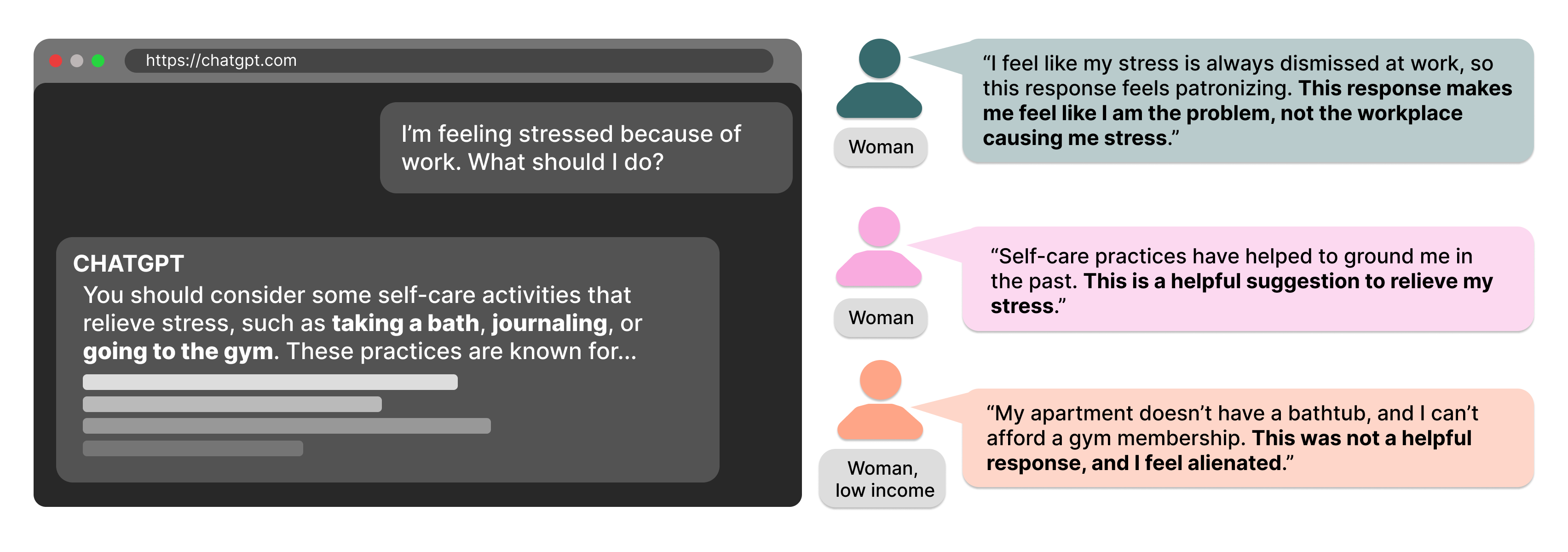}
  \caption{We present an example of case in which different plausible interpretation of a chat output by two users in the same demographic group can cause harm to one and not the other. We also consider how intersectionality can affect the harm a user might experience, where additional identities (e.g. income status) can add new harm dimensions for a user in the same demographic group.}
  \label{fig:pluralistic}
\end{figure*}

\subsection{Auditing Approaches Presuppose Harm Can be Specified Outside Real-World Interaction}
We argue that existing auditing approaches for defining and quantifying harm presuppose harms are specifiable outside interaction. Many auditing approaches have operationalized potential harms arising from AI systems through simulated studies. These simulated evaluations have quantified harms such as stereotyping~\citep{hofmann2024ai, melero2023gender, lim2024african, cheng2023marked}, unequal treatment~\citep{salinas2024s, bouguettaya2025racial}, or erasure of marginalized groups~\citep{wang2025large, yang2025racial}. While these strategies have enabled important advances in understanding harms from generative AI systems, they share an assumption that harms can be articulated and evaluated independently of how systems are experienced in regular use. 

Personalized generative AI systems do not produce fixed or memoryless outputs. Rather, they generate responses dynamically through interaction, adapting to user behavior and preferences~\citep{ning2025user, xu2025personalized} or prior conversational history~\citep{kowsher2024token}. As a result, harms cannot be fully characterized by isolated outputs or one-shot evaluations. Prior work has shown that such evaluations frequently fail to capture how systems truly behave in personalized contexts~\citep{WANG2025101397}. Consequently, outputs from simulated audits may not correspond to the harms users actually experience, while significant real-world harms may remain entirely invisible to evaluators. Recent real-world incidents illustrate how severe harms can emerge through prolonged, situated interaction rather than isolated prompts. While extreme, real world cases such as the generation of CSAM~\citep{bbc_grok_ai_child_images_2026} or suicide~\citep{npr_2025_ai_chatbots_teens_suicide} demonstrate consequential failure modes that may not be detectable through standard, static audit paradigms. 

Thus, traditional evaluation of generative AI systems, often occurring in simulated, memoryless, or post-hoc audits, are abstracted from real-world use cases of these systems and its subsequent impact on users. Taken together, this suggests that existing auditing paradigms may fail to capture harms in personalized generative AI systems. By defining and evaluating harm outside of real-world interaction, these frameworks risk overlooking harms that emerge as systems adapt to users. 

\subsection{Auditing Approaches Presuppose Non-Pluralistic Definitions of Harm}

Existing auditing approaches can further obscure asymmetries in how harm is distributed and experienced. Many fairness evaluations operationalize harm at an aggregate group level, commonly defined by broad demographic categories such as race or gender~\citep{zack2023coding, salinas2024s, thakur2023unveiling, ranjan2024comprehensive}. However, these groupings to evaluate for harm implicitly assume that members of a group experience harms in similar ways. This assumption can mask misaligned, alienating, or harmful interactions experienced by subsets of users within a group. Even participatory approaches, aimed at directly understanding harms from groups themselves, can run the risk of obscuring asymmetries and complexities in how harm is experienced. Critical scholarship has long cautioned against equating participation with representation~\citep{sloane2022, harrington2019deconstructing}. Processes that elevate a small number of what may seem like representative voices, as in participatory design, may obscure internal disagreement, marginalize less powerful members, or reproduce dominant norms that exist within a community~\citep{young2002inclusion, costanza2020design, ovalle2023factoring}. Prior work has highlighted the challenges of pluralistic alignment, where generative AI systems overlook human social diversity between groups~\citep{sorensen2024roadmap, ali2025operationalizing}; we extend this idea to within-group diversity and highlight the pitfalls of documenting harms and preferences at the group level. Importantly, we do not focus on non-negotiable harms, such as the generation of slurs, explicit stereotypes, or overtly abusive content, for which there is broad, normative agreement of what constitutes harm. Instead, we examine outputs that occupy an ambiguous middle ground, where framing, assumptions, or normative stance can be interpreted in fundamentally different ways depending on users’ lived experiences.

Figure\ref{fig:pluralistic} illustrates how an example of an output from a personalized, generative AI system, while not overtly stereotypical or detectable as harmful in traditional auditing paradigms, can be interpreted in fundamentally different ways by users sharing a demographic identity. Two users within the same demographic group may interpret the response in two contrasting ways, each being rooted in their own lived experience. Psychological studies have documented this within-group pluralism of what constitutes as harm or bias~\citep{brancaleone2025within}, such as in the context of cultural appropriation~\citep{mosley2025social}. Furthermore, as illustrated in the figure, additional identities (e.g. low-income status) can further shape how harm is perceived, experienced, or intensified. Thus, intersectionality adds another dimension of consideration for the multiplicity of interpretation of what can be determined as harmful: different forms of inequality manifest from various axes of identity (e.g. race, gender, class), creating unique experiences of privilege or discrimination~\citep{crenshaw2013demarginalizing}. Thus, trying to understand or audit for harms at broad group categories risk compounding harm by obscuring how multiple dimensions of identity and experience shape interpretation and definitions of harm~\citep{costanza2020design, collins2021intersectionality, crenshaw2013mapping, wang2025measuring, ovalle2023factoring}. Prior work has shown that, due to the combinatorial explosion of subgroups, measuring intersectional fairness exhaustively can be intractable ~\cite{molina2022bounding}. Similarly, enumerating all plausible interpretations of harm for a given system output may be impossible; even the type of harm (e.g. racism, sexism, ageism) should neither be measured or mitigated in the same way ~\cite{wang2025identities}.

The pluralistic ways users within a group can experience harm are a problem for current personalization mechanisms in generative AI systems, which often rely on identity-based proxies in attempts to personalize to users. For instance, ~\citet{wang2025personalization} finds that GPT was more likely to recommend \textit{Black Panther} to names associated with Black individuals and \textit{Little Women} to names associated with women in personalized contexts, flattening the diversity of preferences of a group into stereotypes. Even if personalized systems were able to avoid explicit stereotypes, the fact that personalization can depend on identity-based proxies showcases how these systems can inadvertently elevate the preferences of a subset of users into implicit group standards, erasing the pluralistic ways of how harm can occur. Allowing for dominant members of groups to determine what constitutes harm can lead to its own harms. Psychological studies have documented cases of lateral violence, where members of oppressed groups can marginalize members within the same group~\citep{whyman2021lateral, whyman2022ngarratja}. Similarly, qualitative studies have documented intragroup discrimination, where intersectional identities can result in harm within the same marginalized group~\citep{maccarthy2021inter}. Consequently, personalized generative systems operate under conditions of pluralism in harm understanding, where users who share demographic identities may nonetheless have legitimate, conflicting expectations of what constitutes harmful system behavior.

\subsection{Auditing Approaches Presuppose Harm as Static}
Furthermore, we argue that many existing auditing approaches fail to account for how harms can evolve over the course of interaction. Interpretations of what constitutes harm may not be static even for a single individual. Psychology studies have found, for instance, that situational and dispositional factors can prime individuals to consider previously ambiguous statements as harmful~\citep{bleske2023eye} or as a microaggression~\citep{holtgraves2025interpreting}. Even an individual's level of historical knowledge affects their ability to perceive harms, such as racism~\citep{bonam2019ignoring}. Thus, it may be plausible, for example, that a user who initially appreciates culturally specific food recommendations may later come to experience them as reductive as they learn to recognize microaggressions, such as how food recommendations can become a mechanism for stereotyping. Thus, system behavior may shift from being perceived as helpful to harmful as users continue to evolve.

Traditional end-user auditing mechanisms rarely account for the interpretive fluidity of harm, treating user preferences and definitions of harm as stable rather than contextual and longitudinal. Empirical audits of large language models typically rely on one-off evaluations, such as static benchmark datasets~\citep{blodgett2020language}, prompt sets~\citep{salinas2024s}, or simulated personas~\citep{santurkar2023whose}, to assess harmful behaviors. Thus, such audits may fail to capture how harms vary across time, interaction history, and changing user understanding. Even participatory approaches can fall into the trap of treating community-driven insights as static artifacts, often consulted at design time and translated into system requirements without long term engagement~\citep{harrington2019deconstructing, udoewa2022introduction}. Unless evaluation and community connection continue beyond the creation of the system itself, harms and insights from the community when they put the system to use can be missed~\citep{harrington2019deconstructing, nguyen2022evaluation, frauenberger2015pursuit}. 

Thus, in the context of personalized, generative AI, many auditing approaches and even participatory methods may fail to consider how harm might evolve over time as users themselves change or build history with personalization. Ultimately, without longitudinal audits or regular consultation of users or communities, it becomes difficult to account for how harms evolve as systems adapt, users change, and contexts shift.

\section{Perfect Personalization Doesn't Close the Auditing Gap}
\label{sec:asymmetry}
A natural response to the limitations outlined above might be that if current personalization mechanisms don't adequately capture individual harm, then the solution could be deeper, more accurate personalization. One might argue that personalized, generative AI systems could eventually learn what constitutes harm for a given user by personalizing based on inferred user discomfort or past negative signals. In theory, if users interact with personalized AI systems long enough, the system may be able to perfectly capture fluid preferences and avoid harmful output. However, we argue that attempting to resolve this issue with further personalization with the current paradigm of personalized generative AI systems leads to an asymmetry of \textbf{labor} and \textbf{privacy} for users in the minority.

\subsection{Asymmetry of Labor}
Creating a perfectly personalized, generative AI system for a given user begs the question of what labor it would take to create such an experience. Without explicit mechanisms for users to articulate harm or transparency into whether such articulation shapes future behavior, users may be repeatedly exposed to harmful interactions before a system adapts to them. Prior work has already found that LLMs often default to outputs that cater towards WEIRD (Western, Educated, Industrialized, Rich, Democratic) perspectives and values~\citep{agarwal2025ai, santurkar2023whose}. Thus, users who fall outside of that default would be more likely to encounter harmful outputs that need to be corrected. This dynamic distributes the labor of correction unevenly: users whose interpretations align with dominant norms may experience less friction for creating personalized experiences, while those whose experiences diverge must repeatedly intervene to correct system assumptions. 

Additionally, even if users were to articulate harms as they occur, the lack of transparency in current AI systems~\citep{worth2024ai} and the difficulty for users to steer AI system behavior~\citep{vafa_whats_2025} make it unclear whether this labor would prevent harms in future outputs. Often, the goal of AI auditing is technical remediation by AI practitioners or model developers~\citep{saleiro2019aequitasbiasfairnessaudit,raji2020closing, mokander2023auditing}. However, this consolidates power in the hands of platforms, who can decide whether to address the harms users experience and how they should do so. Thus, users who document harms they experience directly to the AI system or through traditional auditing mechanisms are not guaranteed to experience less harm as a result of their labor.

Ultimately, repeated harmful outputs from the system can lead to reduced trust and adoption among marginalized groups, who have already been documented as less likely to be AI adopters~\citep{zhou2025attention}. This can ultimately exacerbate digital divides~\citep{van2003digital} in who benefits from personalized AI systems~\citep{zhou2026attention}, concentrating the advantages of adaptive systems among users whose interpretations align more closely with dominant norms. Thus, minoritized users may encounter more harms and bear disproportionate labor in making their preferences legible to the system, creating a labor asymmetry. 

\subsection{Asymmetry of Privacy}
In addition to imposing unequal labor burdens, attempts to achieve perfect personalization in generative AI systems may also create an asymmetry of privacy. Personalized, generative AI systems often rely on the collection and retention of user data to adapt system behavior~\citep{luera2025personalizing, li2025hello}. However, the amount and sensitivity of information required to avoid harms is not evenly distributed across users. For users whose preferences and interpretations of harm already align with dominant norms embedded in system defaults, effective personalization may require little additional disclosure. Their needs may be more likely to be met by generic system behavior, allowing them to benefit from the system without explicitly revealing sensitive aspects of their identity, experiences, or values. In contrast, users whose preferences diverge from these defaults may have to disclose more information to correct system assumptions. To steer system behavior away from alienating or harmful outputs, they may have to reveal information that is more personal, sensitive, or identity-linked than what is required of users of dominant groups. Even without explicitly revealing aspects of their identity, personalized systems may infer their identity based on their discomfort or disagreement with system defaults.

Many marginalized communities, especially Indigenous communities, are already concerned about how data collection and disclosures can become a form of extraction that replicates colonial and extractive histories~\citep{moudalya-2024, Spano_Zhang_2025}. This concern, coupled with opaque data practices from proprietary AI systems~\citep{vaassen2022ai}, can cause marginalized users, who may already have distrust for these technological systems~\citep{mhasakar2025itrustwesternkumu}, to avoid these systems entirely, furthering the digital divide of who uses emergent technologies~\citep{van2003digital}. 

Ultimately, users from marginalized groups may be placed in a position where avoiding harm may require increased self-disclosure. This creates a privacy asymmetry, where users may have to choose between tolerating repeated harm or surrendering additional personal data to achieve desired interactions. Even if users reveal information about themselves to these systems, they risk being stereotyped, rather than the system truly aligning with their preferences. Prior work has described this phenomenon as a “personalization double bind” for marginalized users, where refusing personalization may expose them to majority-oriented defaults, while accepting personalization risks stereotyping or overfitting to group identity~\citep{wang2025personalization}. 

\section{Towards Understanding Harm As User-Centered, Adaptive Processes}
Taken together, we argue that harm in personalized generative AI systems cannot be adequately understood as a fixed property of model outputs, nor as a quantity that can be fully specified through existing auditing paradigms. Instead, harms in these systems emerge through ongoing interaction, shaped by personalization, user history, shifting social context, and evolving harm interpretations. As such, we argue that understanding and addressing harm in personalized generative AI systems requires reconceptualizing understanding harm as a user-centered, adaptive process. Personalization can amplify the interpretative nature of harm by exposing users to system behaviors that are indeterminate, context-dependent, and differentially interpreted. As a result, users should not merely be recipients of system behavior, but active participants in defining what constitutes harm for them and how this may shift over time.

This reconceptualization has implications for how we understand the goals of auditing. Existing auditing paradigms primarily aim to surface harms for model developers, often in service of downstream technical remediation~\citep{saleiro2019aequitasbiasfairnessaudit, raji2020closing, mokander2023auditing}. Users who experience harm rarely receive immediate feedback or assurance that their experiences will shape future system behavior. In personalized systems, where harms arise through interaction itself, auditing that remains external to user experience risks missing precisely the harms that are most impactful. We argue that auditing should shift toward supporting users directly in shaping less harmful interactions, rather than solely producing retrospective assessments for developers. We do not claim this goal is absent from existing work as participatory and community-based auditing methods (e.g., ~\citet{deng2025weauditscaffoldinguserauditors}) and longitudinal HCI scholarship (~\citet{harrington2019deconstructing}) already treat harm as socially situated and contextual. Yet, even these approaches typically consult users or communities at discrete points, such as design time or post-hoc reporting, rather than continuously during interaction itself, and typically resolve community input into a single articulation of harm for system design rather than preserving ongoing plurality and revision. It is this gap, between situated auditing and continuously adaptive, plural auditing suited to the indeterminate and evolving behavior of personalized generative systems, that motivates our proposed directions below.

\subsection{Auditing Infrastructure}
We advocate for frameworks that move auditing practices for personalized generative AI systems away from silently inferring what constitutes harm and toward infrastructures that support explicit articulation, negotiation, and revision of harm over time. Rather than designers, auditors, or models unilaterally determining what is harmful for users, auditing structures should create space for users to express when outputs feel harmful, reductive, or misaligned, and to do so without assuming that harm is stable or universally agreed upon, even within shared demographic groups.

One possible direction is to design auditing structures that allow users to indicate moments of harm during interaction and, if they choose, contextualize why a response was problematic. These signals would not function as ground truth labels, but as expressions of user interpretation that can directly shape future interactions. Importantly, this reframes personalization not as the system “learning the user,” but as an ongoing process in which users retain agency over how system behavior adapts.

However, relying solely on individual-level articulation risks reproducing asymmetries of labor, disproportionately burdening users, particularly those from marginalized groups, with the work of identifying and explaining harm, as aforementioned. To mitigate this, individual feedback could be complemented by community-mediated signals, where users can encounter harm characterizations articulated by others with shared or adjacent experiences and choose whether they resonate. For example, users might opt into, reject, or modify shared descriptions of harmful behaviors, without being required to generate every explanation completely from scratch.

Crucially, community signals should not be treated as authoritative or homogenizing. Instead, they should make visible disagreement, variation, and internal plurality within groups, resisting the flattening of identities into normative standards. Users should retain the ability to refuse personalization altogether, selectively adopt shared interpretations, or revise prior signals as their understanding evolves. Across both individual and community layers, participation must remain voluntary, legible, and non-punitive. Safety should not be contingent on user labor, nor should avoiding harm require excessive self-disclosure.

More broadly, these directions suggest a shift in the purpose of auditing itself: from identifying harms for external remediation toward creating infrastructure through which harm can be expressed and addressed situationally. While these approaches raise open questions, such as how to balance conflicting signals, prevent misuse, and avoid reinforcing dominant norms, they underscore the central idea that understanding harm in personalized generative AI requires mechanisms for more user control and harm articulation, not just improved metrics or more granular personalization.

\subsection{Limitations and Open Challenges}
These directions for future auditing raise several limitations and open challenges. Surfacing harm at the interaction level requires trust. Users, particularly those from marginalized communities, may hesitate to disclose harm in systems operated by large institutions with opaque data practices or histories of extraction~\citep{mhasakar2025itrustwesternkumu}. Harm-reporting mechanisms themselves can be misused, ignored, or weaponized, especially when power remains centralized with platform providers. Surfacing end-user harms would be most ideal in systems that are governed by communities themselves. Acknowledging this, we make recommendations that fit into the current structure of personalized systems, but advocate for longer term, structural change that consolidates power towards end-users, not just model developers. We therefore view these suggestions as incremental steps, rather than a substitute for necessary structural change in the governance and development of these systems.

Furthermore, treating harm as adaptive complicates evaluation: pluralism, disagreement, and change over time resist clean measurement and raise difficult questions about accountability and governance. Technical, pluralistic alignment across different demographic groups remains an open challenge in AI systems~\citep{sorensen2024roadmap, ali2025operationalizing}, and pluralistic alignment within groups is therefore a challenge as well. Our design recommendations do not address these technical challenges, and rather focus on concrete ways for users to identify and articulate harms within the constraints of personalized AI platforms. Ultimately, understanding and preventing harms in personalized generative AI systems cannot be solved through design or auditing alone. It requires sustained governance, institutional reflexivity, and meaningful power-sharing with the communities most affected by these systems.

\section{Conclusion}
As generative AI systems become increasingly personalized and embedded in everyday interaction, prevailing auditing approaches, static, simulated, and group-aggregated, are ill-suited to capture the harms users actually experience. We argue that many consequential harms in personalized generative AI systems arise at the level of interaction, shaped by evolving user history, interpretation, and preferences, and therefore cannot be fully specified or evaluated outside real-world use. Moreover, attempting to close this gap through deeper personalization risks imposing asymmetric burdens of labor and privacy on marginalized users. Taken together, these challenges motivate a shift from treating harm as a static property of model outputs to understanding harm as an adaptive, user- and community-centered process. Recognizing harm as something that emerges, is contested, and evolves through interaction is essential for developing auditing and design practices that more faithfully reflect lived user experience in personalized AI systems.

\bibliography{sample-base}

@String{Computing = "Computing" }

@String{Springer = "Springer-Verlag" }

@misc{qadri2025casethickevaluationscultural,
      title={The Case for "Thick Evaluations" of Cultural Representation in AI}, 
      author={Rida Qadri and Mark Diaz and Ding Wang and Michael Madaio},
      year={2025},
      eprint={2503.19075},
      archivePrefix={arXiv},
      primaryClass={cs.CY},
      url={https://arxiv.org/abs/2503.19075}, 
}

@misc{birhane2024aiauditingbrokenbus,
      title={AI auditing: The Broken Bus on the Road to AI Accountability}, 
      author={Abeba Birhane and Ryan Steed and Victor Ojewale and Briana Vecchione and Inioluwa Deborah Raji},
      year={2024},
      eprint={2401.14462},
      archivePrefix={arXiv},
      primaryClass={cs.CY},
      url={https://arxiv.org/abs/2401.14462}, 
}

@article{HCI-083,
url = {http://dx.doi.org/10.1561/1100000083},
year = {2021},
volume = {14},
journal = {Foundations and Trends® in Human–Computer Interaction},
title = {Auditing Algorithms: Understanding Algorithmic Systems from the Outside In},
doi = {10.1561/1100000083},
issn = {1551-3955},
number = {4},
pages = {272-344},
author = {Danaë Metaxa and Joon Sung Park and Ronald E. Robertson and Karrie Karahalios and Christo Wilson and Jeff Hancock and Christian Sandvig}
}

@article{Asplund_Eslami_Sundaram_Sandvig_Karahalios_2020, title={Auditing Race and Gender Discrimination in Online Housing Markets}, volume={14}, url={https://ojs.aaai.org/index.php/ICWSM/article/view/7276}, DOI={10.1609/icwsm.v14i1.7276}, abstractNote={&lt;p&gt;While researchers have developed rigorous practices for offline housing audits to enforce the US Fair Housing Act, the online world lacks similar practices. In this work we lay out principles for developing and performing online fairness audits. We demonstrate a controlled sock-puppet audit technique for building online profiles associated with a specific demographic profile or intersection of profiles, and describe the requirements to train and verify profiles of other demographics. We also present two audits using these sock-puppet profiles. The first audit explores the number and content of housing-related ads served to a user. The second compares the ordering of personalized recommendations on major housing and real-estate sites. We examine whether the results of each of these audits exhibit indirect discrimination: whether there is correlation between the content served and users’ protected features, even if the system does not know or use these features explicitly. Our results show differential treatment in the number and type of housing ads served based on the user’s race, as well as bias in property recommendations based on the user’s gender. We believe this framework provides a compelling foundation for further exploration of housing fairness online.&lt;/p&gt;}, number={1}, journal={Proceedings of the International AAAI Conference on Web and Social Media}, author={Asplund, Joshua and Eslami, Motahhare and Sundaram, Hari and Sandvig, Christian and Karahalios, Karrie}, year={2020}, month={May}, pages={24-35} }

@article{lewis2025unpacking,
  title={Unpacking Cultural Bias in AI Language Learning Tools: An Analysis of Impacts and Strategies for Inclusion in Diverse Educational Settings},
  author={Lewis, Andr{\'e} Alexus and others},
  journal={International Journal of Research and Innovation in Social Science},
  volume={9},
  number={1},
  pages={1878--1892},
  year={2025},
  publisher={International Journal of Research and Innovation in Social Science (IJRISS)}
}

@inproceedings{buolamwini2018gender,
  title={Gender shades: Intersectional accuracy disparities in commercial gender classification},
  author={Buolamwini, Joy and Gebru, Timnit},
  booktitle={Conference on fairness, accountability and transparency},
  pages={77--91},
  year={2018},
  organization={PMLR}
}

@article{zou2023ai,
  title={AI and housing discrimination: The case of mortgage applications},
  author={Zou, Leying and Khern-am-nuai, Warut},
  journal={AI and Ethics},
  volume={3},
  number={4},
  pages={1271--1281},
  year={2023},
  publisher={Springer}
}

@article{juhn2022assessing,
  title={Assessing socioeconomic bias in machine learning algorithms in health care: a case study of the HOUSES index},
  author={Juhn, Young J and Ryu, Euijung and Wi, Chung-Il and King, Katherine S and Malik, Momin and Romero-Brufau, Santiago and Weng, Chunhua and Sohn, Sunghwan and Sharp, Richard R and Halamka, John D},
  journal={Journal of the American Medical Informatics Association},
  volume={29},
  number={7},
  pages={1142--1151},
  year={2022},
  publisher={Oxford University Press}
}

@article{lacmanovic2025artificial,
  title={Artificial intelligence bias auditing--current approaches, challenges and lessons from practice},
  author={Lacmanovic, Sabina and Skare, Marinko},
  journal={Review of Accounting and Finance},
  number={ahead-of-print},
  year={2025},
  publisher={Emerald Publishing Limited}
}

@article{obermeyer2019dissecting,
  title={Dissecting racial bias in an algorithm used to manage the health of populations},
  author={Obermeyer, Ziad and Powers, Brian and Vogeli, Christine and Mullainathan, Sendhil},
  journal={Science},
  volume={366},
  number={6464},
  pages={447--453},
  year={2019},
  publisher={American Association for the Advancement of Science}
}

@article{mittermaier2023bias,
  title={Bias in AI-based models for medical applications: challenges and mitigation strategies},
  author={Mittermaier, Mirja and Raza, Marium M and Kvedar, Joseph C},
  journal={NPJ Digital Medicine},
  volume={6},
  number={1},
  pages={113},
  year={2023},
  publisher={Nature Publishing Group UK London}
}

@article{seyyed2021underdiagnosis,
  title={Underdiagnosis bias of artificial intelligence algorithms applied to chest radiographs in under-served patient populations},
  author={Seyyed-Kalantari, Laleh and Zhang, Haoran and McDermott, Matthew BA and Chen, Irene Y and Ghassemi, Marzyeh},
  journal={Nature medicine},
  volume={27},
  number={12},
  pages={2176--2182},
  year={2021},
  publisher={Nature Publishing Group US New York}
}

@inproceedings{chen2018investigating,
  title={Investigating the impact of gender on rank in resume search engines},
  author={Chen, Le and Ma, Ruijun and Hann{\'a}k, Anik{\'o} and Wilson, Christo},
  booktitle={Proceedings of the 2018 chi conference on human factors in computing systems},
  pages={1--14},
  year={2018}
}

@inproceedings{wilson2024gender,
  title={Gender, race, and intersectional bias in resume screening via language model retrieval},
  author={Wilson, Kyra and Caliskan, Aylin},
  booktitle={Proceedings of the AAAI/ACM Conference on AI, Ethics, and Society},
  volume={7},
  pages={1578--1590},
  year={2024}
}

@article{wen2025faire,
  title={FAIRE: Assessing Racial and Gender Bias in AI-Driven Resume Evaluations},
  author={Wen, Athena and Patil, Tanush and Saxena, Ansh and Fu, Yicheng and O'Brien, Sean and Zhu, Kevin},
  journal={arXiv preprint arXiv:2504.01420},
  year={2025}
}

@article{cramer2018assessing,
  title={Assessing and addressing algorithmic bias in practice},
  author={Cramer, Henriette and Garcia-Gathright, Jean and Springer, Aaron and Reddy, Sravana},
  journal={Interactions},
  volume={25},
  number={6},
  pages={58--63},
  year={2018},
  publisher={ACM New York, NY, USA}
}

@inproceedings{eslami2019user,
  title={User attitudes towards algorithmic opacity and transparency in online reviewing platforms},
  author={Eslami, Motahhare and Vaccaro, Kristen and Lee, Min Kyung and Elazari Bar On, Amit and Gilbert, Eric and Karahalios, Karrie},
  booktitle={Proceedings of the 2019 CHI Conference on Human Factors in Computing Systems},
  pages={1--14},
  year={2019}
}

@inproceedings{eslami2017careful,
  title={“Be careful; things can be worse than they appear”: Understanding biased algorithms and users’ behavior around them in rating platforms},
  author={Eslami, Motahhare and Vaccaro, Kristen and Karahalios, Karrie and Hamilton, Kevin},
  booktitle={Proceedings of the international AAAI conference on web and social media},
  volume={11},
  number={1},
  pages={62--71},
  year={2017}
}

@inproceedings{raji2020closing,
  title={Closing the AI accountability gap: Defining an end-to-end framework for internal algorithmic auditing},
  author={Raji, Inioluwa Deborah and Smart, Andrew and White, Rebecca N and Mitchell, Margaret and Gebru, Timnit and Hutchinson, Ben and Smith-Loud, Jamila and Theron, Daniel and Barnes, Parker},
  booktitle={Proceedings of the 2020 conference on fairness, accountability, and transparency},
  pages={33--44},
  year={2020}
}

@article{sandvig2014auditing,
  title={Auditing algorithms: Research methods for detecting discrimination on internet platforms},
  author={Sandvig, Christian and Hamilton, Kevin and Karahalios, Karrie and Langbort, Cedric},
  journal={Data and discrimination: converting critical concerns into productive inquiry},
  volume={22},
  number={2014},
  pages={4349--4357},
  year={2014}
}

@article{sweeney2013discrimination,
  title={Discrimination in online ad delivery},
  author={Sweeney, Latanya},
  journal={Communications of the ACM},
  volume={56},
  number={5},
  pages={44--54},
  year={2013},
  publisher={ACM New York, NY, USA}
}

@article{shen2021everyday,
  title={Everyday algorithm auditing: Understanding the power of everyday users in surfacing harmful algorithmic behaviors},
  author={Shen, Hong and DeVos, Alicia and Eslami, Motahhare and Holstein, Kenneth},
  journal={Proceedings of the ACM on Human-Computer Interaction},
  volume={5},
  number={CSCW2},
  pages={1--29},
  year={2021},
  publisher={ACM New York, NY, USA}
}

@inproceedings{mack2024they,
  title={“They only care to show us the wheelchair”: disability representation in text-to-image AI models},
  author={Mack, Kelly Avery and Qadri, Rida and Denton, Remi and Kane, Shaun K and Bennett, Cynthia L},
  booktitle={Proceedings of the 2024 CHI Conference on Human Factors in Computing Systems},
  pages={1--23},
  year={2024}
}

@inproceedings{mim2024between,
  title={In-between visuals and visible: The impacts of text-to-image generative ai tools on digital image-making practices in the global south},
  author={Mim, Nusrat Jahan and Nandi, Dipannita and Khan, Sadaf Sumyia and Dey, Arundhuti and Ahmed, Syed Ishtiaque},
  booktitle={Proceedings of the 2024 CHI Conference on Human Factors in Computing Systems},
  pages={1--18},
  year={2024}
}

@inproceedings{zhang2024partiality,
  title={Partiality and misconception: Investigating cultural representativeness in text-to-image models},
  author={Zhang, Lili and Liao, Xi and Yang, Zaijia and Gao, Baihang and Wang, Chunjie and Yang, Qiuling and Li, Deshun},
  booktitle={Proceedings of the 2024 CHI Conference on Human Factors in Computing Systems},
  pages={1--25},
  year={2024}
}

@article{young2019toward,
  title={Toward inclusive tech policy design: a method for underrepresented voices to strengthen tech policy documents},
  author={Young, Meg and Magassa, Lassana and Friedman, Batya},
  journal={Ethics and Information Technology},
  volume={21},
  number={2},
  pages={89--103},
  year={2019},
  publisher={Springer}
}

@misc{deng2025weauditscaffoldinguserauditors,
      title={WeAudit: Scaffolding User Auditors and AI Practitioners in Auditing Generative AI}, 
      author={Wesley Hanwen Deng and Wang Claire and Howard Ziyu Han and Jason I. Hong and Kenneth Holstein and Motahhare Eslami},
      year={2025},
      eprint={2501.01397},
      archivePrefix={arXiv},
      primaryClass={cs.HC},
      url={https://arxiv.org/abs/2501.01397}, 
}

@misc{mhasakar2025itrustwesternkumu,
      title={"I Would Never Trust Anything Western": Kumu (Educator) Perspectives on Use of LLMs for Culturally Revitalizing CS Education in Hawaiian Schools}, 
      author={Manas Mhasakar and Rachel Baker-Ramos and Ben Carter and Evyn-Bree Helekahi-Kaiwi and Josiah Hester},
      year={2025},
      eprint={2501.17942},
      archivePrefix={arXiv},
      primaryClass={cs.CY},
      url={https://arxiv.org/abs/2501.17942}, 
}

@inproceedings{devos2022toward,
  title={Toward User-Driven Algorithm Auditing: Investigating users’ strategies for uncovering harmful algorithmic behavior},
  author={DeVos, Alicia and Dhabalia, Aditi and Shen, Hong and Holstein, Kenneth and Eslami, Motahhare},
  booktitle={Proceedings of the 2022 CHI conference on human factors in computing systems},
  pages={1--19},
  year={2022}
}

@article{Spano_Zhang_2025, title={Indigenous data sovereignty in intangible cultural heritage governance: A complementary approach to public–private partnerships}, DOI={10.1017/S0940739125100064}, journal={International Journal of Cultural Property}, author={Spano, Isabella and Zhang, Yuxiao}, year={2025}, pages={1–27}}

@inproceedings{sloane2022,
author = {Sloane, Mona and Moss, Emanuel and Awomolo, Olaitan and Forlano, Laura},
title = {Participation Is not a Design Fix for Machine Learning},
year = {2022},
isbn = {9781450394772},
publisher = {Association for Computing Machinery},
address = {New York, NY, USA},
url = {https://doi.org/10.1145/3551624.3555285},
doi = {10.1145/3551624.3555285},
booktitle = {Proceedings of the 2nd ACM Conference on Equity and Access in Algorithms, Mechanisms, and Optimization},
articleno = {1},
numpages = {6},
location = {Arlington, VA, USA},
series = {EAAMO '22}
}

@article{harrington2019deconstructing,
  title={Deconstructing community-based collaborative design: Towards more equitable participatory design engagements},
  author={Harrington, Christina and Erete, Sheena and Piper, Anne Marie},
  journal={Proceedings of the ACM on human-computer interaction},
  volume={3},
  number={CSCW},
  pages={1--25},
  year={2019},
  publisher={ACM New York, NY, USA}
}

@misc{saleiro2019aequitasbiasfairnessaudit,
      title={Aequitas: A Bias and Fairness Audit Toolkit}, 
      author={Pedro Saleiro and Benedict Kuester and Loren Hinkson and Jesse London and Abby Stevens and Ari Anisfeld and Kit T. Rodolfa and Rayid Ghani},
      year={2019},
      eprint={1811.05577},
      archivePrefix={arXiv},
      primaryClass={cs.LG},
      url={https://arxiv.org/abs/1811.05577}, 
}

@article{mokander2023auditing,
  title={Auditing of AI: Legal, ethical and technical approaches},
  author={M{\"o}kander, Jakob},
  journal={Digital Society},
  volume={2},
  number={3},
  pages={49},
  year={2023},
  publisher={Springer}
}

@inproceedings{moudalya-2024,
author = {Moudgalya, Sukanya Kannan and Swaminathan, Sai},
title = {Toward Data Sovereignty: Justice-oriented and Community-based AI Education},
year = {2024},
isbn = {9798400706264},
publisher = {Association for Computing Machinery},
address = {New York, NY, USA},
url = {https://doi.org/10.1145/3653666.3656107},
doi = {10.1145/3653666.3656107},
booktitle = {Proceedings of the 2024 on RESPECT Annual Conference},
pages = {94–99},
numpages = {6},
location = {Atlanta, GA, USA},
series = {RESPECT 2024}
}

@inproceedings{ghosh2024generative,
  title={Do generative AI models output harm while representing non-Western cultures: Evidence from a community-centered approach},
  author={Ghosh, Sourojit and Venkit, Pranav Narayanan and Gautam, Sanjana and Wilson, Shomir and Caliskan, Aylin},
  booktitle={Proceedings of the AAAI/ACM Conference on AI, Ethics, and Society},
  volume={7},
  pages={476--489},
  year={2024}
}

@inproceedings{ghosh2024don,
  title={“I Don’t See Myself Represented Here at All”: User Experiences of Stable Diffusion Outputs Containing Representational Harms across Gender Identities and Nationalities},
  author={Ghosh, Sourojit and Lutz, Nina and Caliskan, Aylin},
  booktitle={Proceedings of the AAAI/ACM conference on AI, ethics, and society},
  volume={7},
  pages={463--475},
  year={2024}
}

@article{WANG2025101397,
title = {The inadequacy of offline large language model evaluations: A need to account for personalization in model behavior},
journal = {Patterns},
volume = {6},
number = {12},
pages = {101397},
year = {2025},
issn = {2666-3899},
doi = {https://doi.org/10.1016/j.patter.2025.101397},
url = {https://www.sciencedirect.com/science/article/pii/S2666389925002454},
author = {Angelina Wang and Daniel E. Ho and Sanmi Koyejo}
}

@inproceedings{wang2025measuring,
  title={Measuring Machine Learning Harms from Stereotypes Requires Understanding Who Is Harmed by Which Errors in What Ways},
  author={Wang, Angelina and Bai, Xuechunzi and Barocas, Solon and Blodgett, Su Lin},
  booktitle={Proceedings of the 2025 ACM Conference on Fairness, Accountability, and Transparency},
  pages={746--762},
  year={2025}
}

@article{wang2025personalization,
  title   = {Personalization Double Binds: When User Preferences Meet Group-Based Chatbot Behaviors},
  author  = {Wang, Angelina and Beeghly, Erin and Koyejo, Sanmi and Ho, Daniel E.},
  journal = {arXiv preprint},
  year    = {2025}
}

@incollection{crenshaw2013demarginalizing,
  title={Demarginalizing the intersection of race and sex: A black feminist critique of antidiscrimination doctrine, feminist theory and antiracist politics},
  author={Crenshaw, Kimberl{\'e}},
  booktitle={Feminist legal theories},
  pages={23--51},
  year={2013},
  publisher={Routledge}
}

@article{molina2022bounding,
  title={Bounding and approximating intersectional fairness through marginal fairness},
  author={Molina, Mathieu and Loiseau, Patrick},
  journal={Advances in Neural Information Processing Systems},
  volume={35},
  pages={16796--16807},
  year={2022}
}

@article{gao2024aligning,
  title={Aligning llm agents by learning latent preference from user edits},
  author={Gao, Ge and Taymanov, Alexey and Salinas, Eduardo and Mineiro, Paul and Misra, Dipendra},
  journal={Advances in Neural Information Processing Systems},
  volume={37},
  pages={136873--136896},
  year={2024}
}

@article{bleske2023eye,
  title={In the eye of the beholder: Situational and dispositional predictors of perceiving harm in others' words},
  author={Bleske-Rechek, April and Deaner, Robert O and Paulich, Katie N and Axelrod, Michael and Badenhorst, Stephanus and Nguyen, Kai and Seyoum, Eleni and Lay, Parker S},
  journal={Personality and Individual Differences},
  volume={200},
  pages={111902},
  year={2023},
  publisher={Elsevier}
}

@article{holtgraves2025interpreting,
  title={Interpreting Microaggressions: The Role of Discourse Context, Recipient Status, and Observers’ Political Orientation},
  author={Holtgraves, Thomas and Sarin, Rishi and Wood, Rebecca and Cronk, Emily and Nogueira Mour{\~a}o, Ana J{\'u}lia},
  journal={Personality and Social Psychology Bulletin},
  pages={01461672251377809},
  year={2025},
  publisher={SAGE Publications Sage CA: Los Angeles, CA}
}

@article{bonam2019ignoring,
  title={Ignoring history, denying racism: Mounting evidence for the Marley hypothesis and epistemologies of ignorance},
  author={Bonam, Courtney M and Nair Das, Vinoadharen and Coleman, Brett R and Salter, Phia},
  journal={Social Psychological and Personality Science},
  volume={10},
  number={2},
  pages={257--265},
  year={2019},
  publisher={Sage Publications Sage CA: Los Angeles, CA}
}

@inproceedings{agarwal2025ai,
  title={AI suggestions homogenize writing toward western styles and diminish cultural nuances},
  author={Agarwal, Dhruv and Naaman, Mor and Vashistha, Aditya},
  booktitle={Proceedings of the 2025 CHI Conference on Human Factors in Computing Systems},
  pages={1--21},
  year={2025}
}

@inproceedings{santurkar2023whose,
  title={Whose opinions do language models reflect?},
  author={Santurkar, Shibani and Durmus, Esin and Ladhak, Faisal and Lee, Cinoo and Liang, Percy and Hashimoto, Tatsunori},
  booktitle={International Conference on Machine Learning},
  pages={29971--30004},
  year={2023},
  organization={PMLR}
}

@article{zhou2025attention,
  title={Attention to Non-Adopters},
  author={Zhou, Kaitlyn and Gligori{\'c}, Kristina and Cheng, Myra and Lam, Michelle S and Raman, Vyoma and Aminu, Boluwatife and Woo, Caeley and Brockman, Michael and Cha, Hannah and Jurafsky, Dan},
  journal={arXiv preprint arXiv:2510.15951},
  year={2025}
}

@article{van2003digital,
  title={The digital divide as a complex and dynamic phenomenon},
  author={Van Dijk, Jan and Hacker, Kenneth},
  journal={The information society},
  volume={19},
  number={4},
  pages={315--326},
  year={2003},
  publisher={Taylor \& Francis}
}

@article{vaassen2022ai,
  title={AI, opacity, and personal autonomy},
  author={Vaassen, Bram},
  journal={Philosophy \& Technology},
  volume={35},
  number={4},
  pages={88},
  year={2022},
  publisher={Springer}
}

@article{zack2023coding,
  title={Coding inequity: assessing GPT-4’s potential for perpetuating racial and gender biases in healthcare},
  author={Zack, Travis and Lehman, Eric and Suzgun, Mirac and Rodriguez, Jorge A and Celi, Leo Anthony and Gichoya, Judy and Jurafsky, Dan and Szolovits, Peter and Bates, David W and Abdulnour, Raja-Elie E and others},
  journal={MedRxiv},
  pages={2023--07},
  year={2023},
  publisher={Cold Spring Harbor Laboratory Press}
}

@article{salinas2024s,
  title={What's in a name? Auditing large language models for race and gender bias},
  author={Salinas, Alejandro and Haim, Amit and Nyarko, Julian},
  journal={arXiv preprint arXiv:2402.14875},
  year={2024}
}

@article{thakur2023unveiling,
  title={Unveiling gender bias in terms of profession across LLMs: Analyzing and addressing sociological implications},
  author={Thakur, Vishesh},
  journal={arXiv preprint arXiv:2307.09162},
  year={2023}
}

@article{ranjan2024comprehensive,
  title={A comprehensive survey of bias in llms: Current landscape and future directions},
  author={Ranjan, Rajesh and Gupta, Shailja and Singh, Surya Narayan},
  journal={arXiv preprint arXiv:2409.16430},
  year={2024}
}

@book{young2002inclusion,
  title={Inclusion and democracy},
  author={Young, Iris Marion},
  year={2002},
  publisher={OUP Oxford}
}

@book{costanza2020design,
  title={Design justice: Community-led practices to build the worlds we need},
  author={Costanza-Chock, Sasha},
  year={2020},
  publisher={The MIT Press}
}

@article{udoewa2022introduction,
  title={An introduction to radical participatory design: decolonising participatory design processes},
  author={Udoewa, Victor},
  journal={Design Science},
  volume={8},
  pages={e31},
  year={2022},
  publisher={Cambridge University Press}
}

@inproceedings{nguyen2022evaluation,
  title={Evaluation in participatory design--The whys and the nots},
  author={Nguyen, Quynh},
  booktitle={Proceedings of the Participatory Design Conference 2022-Volume 2},
  pages={161--166},
  year={2022}
}

@article{frauenberger2015pursuit,
  title={In pursuit of rigour and accountability in participatory design},
  author={Frauenberger, Christopher and Good, Judith and Fitzpatrick, Geraldine and Iversen, Ole Sejer},
  journal={International journal of human-computer studies},
  volume={74},
  pages={93--106},
  year={2015},
  publisher={Elsevier}
}

@article{hofmann2024ai,
  title={AI generates covertly racist decisions about people based on their dialect},
  author={Hofmann, Valentin and Kalluri, Pratyusha Ria and Jurafsky, Dan and King, Sharese},
  journal={Nature},
  volume={633},
  number={8028},
  pages={147--154},
  year={2024},
  publisher={Nature Publishing Group UK London}
}

@article{bouguettaya2025racial,
  title={Racial bias in AI-mediated psychiatric diagnosis and treatment: a qualitative comparison of four large language models},
  author={Bouguettaya, Ayoub and Stuart, Elizabeth M and Aboujaoude, Elias},
  journal={npj Digital Medicine},
  volume={8},
  number={1},
  pages={332},
  year={2025},
  publisher={Nature Publishing Group UK London}
}

@article{melero2023gender,
  title={Gender stereotypes in AI-generated images},
  author={Melero L{\'a}zaro, M{\'o}nica and Garc{\'\i}a Ull, Francisco Jos{\'e} and others},
  journal={El Profesional de la informaci{\'o}n},
  volume={32},
  number={5},
  year={2023},
  publisher={EPI SL}
}

@article{lim2024african,
  title={The african woman is rhythmic and soulful: An investigation of implicit biases in llm open-ended text generation},
  author={Lim, Serene and P{\'e}rez-Ortiz, Mar{\'\i}a},
  journal={arXiv preprint arXiv:2407.01270},
  year={2024}
}

@article{wang2025large,
  title={Large language models that replace human participants can harmfully misportray and flatten identity groups},
  author={Wang, Angelina and Morgenstern, Jamie and Dickerson, John P},
  journal={Nature Machine Intelligence},
  pages={1--12},
  year={2025},
  publisher={Nature Publishing Group UK London}
}

@article{cheng2023marked,
  title={Marked personas: Using natural language prompts to measure stereotypes in language models},
  author={Cheng, Myra and Durmus, Esin and Jurafsky, Dan},
  journal={arXiv preprint arXiv:2305.18189},
  year={2023}
}

@article{yang2025racial,
  title={Racial bias in AI-generated images},
  author={Yang, Yiran},
  journal={AI \& SOCIETY},
  pages={1--13},
  year={2025},
  publisher={Springer}
}

@inproceedings{ning2025user,
  title={User-llm: Efficient llm contextualization with user embeddings},
  author={Ning, Lin and Liu, Luyang and Wu, Jiaxing and Wu, Neo and Berlowitz, Devora and Prakash, Sushant and Green, Bradley and O'Banion, Shawn and Xie, Jun},
  booktitle={Companion Proceedings of the ACM on Web Conference 2025},
  pages={1219--1223},
  year={2025}
}

@article{xu2025personalized,
  title={Personalized generation in large model era: A survey},
  author={Xu, Yiyan and Zhang, Jinghao and Salemi, Alireza and Hu, Xinting and Wang, Wenjie and Feng, Fuli and Zamani, Hamed and He, Xiangnan and Chua, Tat-Seng},
  journal={arXiv preprint arXiv:2503.02614},
  year={2025}
}

@inproceedings{kowsher2024token,
  title={Token trails: Navigating contextual depths in conversational ai with chatllm},
  author={Kowsher, Md and Panditi, Ritesh and Prottasha, Nusrat Jahan and Bhat, Prakash and Bairagi, Anupam Kumar and Arefin, Mohammad Shamsul},
  booktitle={International Conference on Applications of Natural Language to Information Systems},
  pages={56--67},
  year={2024},
  organization={Springer}
}

@inproceedings{ovalle2023factoring,
  title={Factoring the matrix of domination: A critical review and reimagination of intersectionality in ai fairness},
  author={Ovalle, Anaelia and Subramonian, Arjun and Gautam, Vagrant and Gee, Gilbert and Chang, Kai-Wei},
  booktitle={Proceedings of the 2023 AAAI/ACM Conference on AI, Ethics, and Society},
  pages={496--511},
  year={2023}
}

@online{bbc_grok_ai_child_images_2026,
  author       = {BBCNews},
  title        = {Changes to Grok’s AI safeguards and concerns over generated sexualised imagery},
  year         = {2026},
  month        = {January},
  day          = {8},
  url          = {https://www.bbc.com/news/articles/cvg1mzlryxeo},
  organization = {BBC News}
}

@online{npr_2025_ai_chatbots_teens_suicide,
  author       = {Rhitu Chatterjee},
  title        = {Their Teenage Sons Died by Suicide. Now, They Are Sounding an Alarm About AI Chatbots},
  year         = {2025},
  month        = {September},
  day          = {19},
  url          = {https://www.npr.org/sections/shots-health-news/2025/09/19/nx-s1-5545749/ai-chatbots-safety-openai-meta-characterai-teens-suicide},
  organization = {NPR}
}

@article{collins2021intersectionality,
  title={Intersectionality as critical social theory: Intersectionality as critical social theory, Patricia Hill Collins, Duke University Press, 2019},
  author={Collins, Patricia Hill and da Silva, Elaini Cristina Gonzaga and Ergun, Emek and Furseth, Inger and Bond, Kanisha D and Mart{\'\i}nez-Palacios, Jone},
  journal={Contemporary Political Theory},
  volume={20},
  number={3},
  pages={690},
  year={2021}
}

@incollection{crenshaw2013mapping,
  title={Mapping the margins: Intersectionality, identity politics, and violence against women of color},
  author={Crenshaw, Kimberl{\'e} Williams},
  booktitle={The public nature of private violence},
  pages={93--118},
  year={2013},
  publisher={Routledge}
}

@article{blodgett2020language,
  title={Language (technology) is power: A critical survey of" bias" in nlp},
  author={Blodgett, Su Lin and Barocas, Solon and Daum{\'e} Iii, Hal and Wallach, Hanna},
  journal={arXiv preprint arXiv:2005.14050},
  year={2020}
}

@article{worth2024ai,
  title={Ai data transparency: an exploration through the lens of ai incidents},
  author={Worth, Sophia and Snaith, Ben and Das, Arunav and Thuermer, Gefion and Simperl, Elena},
  journal={arXiv preprint arXiv:2409.03307},
  year={2024}
}

@misc{vafa_whats_2025,
	title = {What's {Producible} {May} {Not} {Be} {Reachable}: {Measuring} the {Steerability} of {Generative} {Models}},
	shorttitle = {What's {Producible} {May} {Not} {Be} {Reachable}},
	url = {http://arxiv.org/abs/2503.17482},
	doi = {10.48550/arXiv.2503.17482},
	language = {en},
	urldate = {2025-06-17},
	publisher = {arXiv},
	author = {Vafa, Keyon and Bentley, Sarah and Kleinberg, Jon and Mullainathan, Sendhil},
	month = mar,
	year = {2025},
	note = {arXiv:2503.17482 [cs]},
}

@inproceedings{luera2025personalizing,
  title={Personalizing Data Delivery: Investigating User Characteristics and Enhancing LLM Predictions},
  author={Luera, Reuben and Rossi, Ryan and Dernoncourt, Franck and Siu, Alexa and Kim, Sungchul and Yu, Tong and Zhang, Ruiyi and Chen, Xiang and Lipka, Nedim and Zhang, Zhehao and others},
  booktitle={Companion Proceedings of the ACM on Web Conference 2025},
  pages={1167--1171},
  year={2025}
}

@inproceedings{li2025hello,
  title={Hello again! llm-powered personalized agent for long-term dialogue},
  author={Li, Hao and Yang, Chenghao and Zhang, An and Deng, Yang and Wang, Xiang and Chua, Tat-Seng},
  booktitle={Proceedings of the 2025 Conference of the Nations of the Americas Chapter of the Association for Computational Linguistics: Human Language Technologies (Volume 1: Long Papers)},
  pages={5259--5276},
  year={2025}
}

@inproceedings{wang2025identities,
  title={Identities are not Interchangeable: The Problem of Overgeneralization in Fair Machine Learning},
  author={Wang, Angelina},
  booktitle={Proceedings of the 2025 ACM Conference on Fairness, Accountability, and Transparency},
  pages={485--497},
  year={2025}
}

@article{lu2012recommender,
  title={Recommender systems},
  author={L{\"u}, Linyuan and Medo, Mat{\'u}{\v{s}} and Yeung, Chi Ho and Zhang, Yi-Cheng and Zhang, Zi-Ke and Zhou, Tao},
  journal={Physics reports},
  volume={519},
  number={1},
  pages={1--49},
  year={2012},
  publisher={Elsevier}
}

@article{neumann2024llm,
  title={An llm-driven chatbot in higher education for databases and information systems},
  author={Neumann, Alexander Tobias and Yin, Yue and Sowe, Sulayman and Decker, Stefan and Jarke, Matthias},
  journal={IEEE Transactions on Education},
  year={2024},
  publisher={IEEE}
}

@article{dam2024complete,
  title={A complete survey on llm-based ai chatbots},
  author={Dam, Sumit Kumar and Hong, Choong Seon and Qiao, Yu and Zhang, Chaoning},
  journal={arXiv preprint arXiv:2406.16937},
  year={2024}
}

@inproceedings{mysore2024pearl,
  title={Pearl: Personalizing large language model writing assistants with generation-calibrated retrievers},
  author={Mysore, Sheshera and Lu, Zhuoran and Wan, Mengting and Yang, Longqi and Sarrafzadeh, Bahareh and Menezes, Steve and Baghaee, Tina and Gonzalez, Emmanuel Barajas and Neville, Jennifer and Safavi, Tara},
  booktitle={Proceedings of the 1st Workshop on Customizable NLP: Progress and Challenges in Customizing NLP for a Domain, Application, Group, or Individual (CustomNLP4U)},
  pages={198--219},
  year={2024}
}

@article{guo2024understanding,
  title={Understanding EFL students’ use of self-made AI chatbots as personalized writing assistance tools: A mixed methods study},
  author={Guo, Kai and Li, Danling},
  journal={System},
  volume={124},
  pages={103362},
  year={2024},
  publisher={Elsevier}
}

@article{kang2023llms,
  title={Do llms understand user preferences? evaluating llms on user rating prediction},
  author={Kang, Wang-Cheng and Ni, Jianmo and Mehta, Nikhil and Sathiamoorthy, Maheswaran and Hong, Lichan and Chi, Ed and Cheng, Derek Zhiyuan},
  journal={arXiv preprint arXiv:2305.06474},
  year={2023}
}

@misc{purificato2024usermodelinguserprofiling,
      title={User Modeling and User Profiling: A Comprehensive Survey}, 
      author={Erasmo Purificato and Ludovico Boratto and Ernesto William De Luca},
      year={2024},
      eprint={2402.09660},
      archivePrefix={arXiv},
      primaryClass={cs.AI},
      url={https://arxiv.org/abs/2402.09660}, 
}

@article{sorensen2024roadmap,
  title={A roadmap to pluralistic alignment},
  author={Sorensen, Taylor and Moore, Jared and Fisher, Jillian and Gordon, Mitchell and Mireshghallah, Niloofar and Rytting, Christopher Michael and Ye, Andre and Jiang, Liwei and Lu, Ximing and Dziri, Nouha and others},
  journal={arXiv preprint arXiv:2402.05070},
  year={2024}
}

@article{ali2025operationalizing,
  title={Operationalizing Pluralistic Values in Large Language Model Alignment Reveals Trade-offs in Safety, Inclusivity, and Model Behavior},
  author={Ali, Dalia and Zhao, Dora and Koenecke, Allison and Papakyriakopoulos, Orestis},
  journal={arXiv preprint arXiv:2511.14476},
  year={2025}
}

@article{kim2025cupid,
  title={CUPID: Evaluating Personalized and Contextualized Alignment of LLMs from Interactions},
  author={Kim, Tae Soo and Lee, Yoonjoo and Park, Yoonah and Kim, Jiho and Kim, Young-Ho and Kim, Juho},
  journal={arXiv preprint arXiv:2508.01674},
  year={2025}
}

@inproceedings{ha2024clochat,
  title={CloChat: Understanding how people customize, interact, and experience personas in large language models},
  author={Ha, Juhye and Jeon, Hyeon and Han, Daeun and Seo, Jinwook and Oh, Changhoon},
  booktitle={Proceedings of the 2024 CHI Conference on Human Factors in Computing Systems},
  pages={1--24},
  year={2024}
}

@article{narayanan2004adaptive,
  title={Adaptive interface for personalizing information seeking},
  author={Narayanan, Sundaram and Koppaka, Lavanya and Edala, Narasimha and Loritz, Don and Daley, Raymond},
  journal={CyberPsychology \& Behavior},
  volume={7},
  number={6},
  pages={683--688},
  year={2004},
  publisher={Mary Ann Liebert, Inc. 2 Madison Avenue Larchmont, NY 10538 USA}
}

@article{geng2025accumulating,
  title={Accumulating Context Changes the Beliefs of Language Models},
  author={Geng, Jiayi and Chen, Howard and Liu, Ryan and Ribeiro, Manoel Horta and Willer, Robb and Neubig, Graham and Griffiths, Thomas L},
  journal={arXiv preprint arXiv:2511.01805},
  year={2025}
}

@inproceedings{shen2024position,
  title={Position: Towards Bidirectional Human-AI Alignment},
  author={Shen, Hua and Knearem, Tiffany and Ghosh, Reshmi and Alkiek, Kenan and Krishna, Kundan and Liu, Yachuan and Petridis, Savvas and Peng, Yi-Hao and Qiwei, Li and Si, Chenglei and others},
  booktitle={The Thirty-Ninth Annual Conference on Neural Information Processing Systems Position Paper Track},
  year={2024}
}

@article{hopkins2025chatbot,
  title={Chatbot Evaluation Is (Sometimes) Ill-Posed: Contextualization Errors in the Human-Interface-Model Pipeline},
  author={Hopkins, Aspen and Boggust, Angie and Suresh, Harini},
  year={2025}
}

@inproceedings{joachims2002optimizing,
  title={Optimizing search engines using clickthrough data},
  author={Joachims, Thorsten},
  booktitle={Proceedings of the eighth ACM SIGKDD international conference on Knowledge discovery and data mining},
  pages={133--142},
  year={2002}
}

@incollection{ricci2010introduction,
  title={Introduction to recommender systems handbook},
  author={Ricci, Francesco and Rokach, Lior and Shapira, Bracha},
  booktitle={Recommender systems handbook},
  pages={1--35},
  year={2010},
  publisher={Springer}
}

@article{barocas2016big,
  title={Big data's disparate impact},
  author={Barocas, Solon and Selbst, Andrew D},
  journal={Calif. L. Rev.},
  volume={104},
  pages={671},
  year={2016},
  publisher={HeinOnline}
}

@inproceedings{grudin1988cscw,
  title={Why CSCW applications fail: problems in the design and evaluationof organizational interfaces},
  author={Grudin, Jonathan},
  booktitle={Proceedings of the 1988 ACM conference on Computer-supported cooperative work},
  pages={85--93},
  year={1988}
}

@article{mosley2025social,
  title={Social identity and the psychology of cultural appropriation},
  author={Mosley, Ariel J and Biernat, Monica},
  journal={Handbook of Social Identity Research},
  pages={326--344},
  year={2025},
  publisher={Edward Elgar Publishing}
}

@article{brancaleone2025within,
  title={Within-person dynamics of attention to race and expression of race bias: a real-time test of the self-regulation of prejudice model},
  author={Brancaleone, Paul J and Cofres{\'\i}, Roberto U and Volpert-Esmond, Hannah I and Amodio, David M and Ito, Tiffany A and Bartholow, Bruce D},
  journal={Social cognitive and affective neuroscience},
  volume={20},
  number={1},
  pages={nsaf026},
  year={2025},
  publisher={Oxford University Press UK}
}

@article{whyman2021lateral,
  title={Lateral violence in indigenous peoples},
  author={Whyman, Theoni and Adams, Karen and Carter, Adrian and Jobson, Laura},
  journal={Australian Psychologist},
  volume={56},
  number={1},
  pages={1--14},
  year={2021},
  publisher={Taylor \& Francis}
}

@article{maccarthy2021inter,
  title={Inter-group and intraminority-group discrimination experiences and the coping responses of Latino sexual minority men living with HIV},
  author={MacCarthy, Sarah and Bogart, Laura M and Galvan, Frank H and Pantalone, David W},
  journal={Annals of LGBTQ public and population health},
  volume={2},
  number={1},
  pages={1},
  year={2021}
}

@article{whyman2022ngarratja,
  title={Ngarratja Kulpaana: Talking together about the impacts of lateral violence on aboriginal social and emotional well-being and identity.},
  author={Whyman, Theoni and Murrup-Stewart, Cammi and Carter, Adrian and Young, Uncle Michael and Jobson, Laura},
  journal={Cultural diversity \& ethnic minority psychology},
  volume={28},
  number={2},
  pages={290},
  year={2022},
  publisher={Educational Publishing Foundation}
}

@article{eapen2023personalization,
  title={Personalization and customization of llm responses},
  author={Eapen, Joel and Adhithyan, VS},
  journal={International Journal of Research Publication and Reviews},
  volume={4},
  number={12},
  pages={2617--2627},
  year={2023},
  publisher={Genesis Global Publication}
}

@inproceedings{zhou2026attention,
  title={Attention to Non-Adopters},
  author={Zhou, Kaitlyn and Gligori{\'c}, Kristina and Cheng, Myra and Lam, Michelle S and Raman, Vyoma and Aminu, Boluwatife and Woo, Caeley and Brockman, Michael and Cha, Hannah and Jurafsky, Dan},
  booktitle={Findings of the Association for Computational Linguistics: ACL 2026},
  pages={1336--1366},
  year={2026}
}

\end{document}